\documentclass[12pt]{iopart}
\usepackage{hyperref}
\usepackage{listings}
\usepackage{graphicx}
\expandafter\let\csname equation*\endcsname\relax
\expandafter\let\csname endequation*\endcsname\relax
\usepackage{amsmath}
\usepackage[sorting=none]{biblatex}
\bibliography{papers}
\usepackage{enumitem}
\usepackage[separate-uncertainty=true]{siunitx}
\usepackage{verbatim}

\begin{document}
\title{Thermodynamics education for energy transformation: a Stirling Engine experiment}

\author{Will Yeadon and Mark Quinn}

\address{Department of Physics and Astronomy, University of Sheffield, Hounsfield Road, Sheffield S3 7RH, UK}
\ead{will.yeadon@sheffield.ac.uk}
\vspace{10pt}
\begin{indented}
\item[]{April 2021}
\end{indented}

\begin{abstract}
We present a thermodynamics experiment suitable for first year undergraduate students employing Stirling Engines to create a demonstration of energy transformation and to measure the mechanical efficiency of such engines. Using an inexpensive transparent chambered Stirling Engine, students can connect concepts such as the theoretical pressure-volume diagram with the physical movements of the engine's pistons and the resultant useful output work of a spinning wheel. We found the majority of students successfully complete this experiment obtaining results similar to when performed by the authors. In addition to the core thermodynamics lesson, this experiment incorporates DC circuits, oscilloscopes, and data analysis so it can be integrated into a wider undergraduate physics course to combine the teaching of multiple subjects.  
\end{abstract}

\begin{section}{Introduction}\label{intro}
Thermodynamics is a key topic in a contemporary physics degree. Yet core concepts such as heat and work are often conflated by students \cite{meltzer-04, vanRoon-94}. Particularly for process functions such as work done through pressure and volume changes of a gas, students may have more difficulty accurately describing the mathematics within a physics context \cite{p-v-pollock}. Conversely a classical mechanics view of mechanical work being the product of force and distance is typically introduced to students before university. Novel mental models and theoretical justifications as methods for teaching thermodynamic concepts has been the subject of much recent research \cite{sunray-concept, breath-engine, energy-entropy-concept}. Providing an experiment for students to perform in addition to a theoretical justification enables the conceptualization of energy transformation to be "anchored".  

In a review article, Mulop \emph{et al.} \cite{review-mulop} highlighted the difficulty students have in visualising thermodynamic concepts as a barrier to learning. Stated difficulties included applying concepts of thermodynamic processes to a real life power plant operation and theories understood as abstractions that have no real life application. Further, students typically can hold varied conceptions of energy transformation that are applied in different situations: the energy transformation of a ball rolling down a slope is conceptually different to the energy transformation involved in the forming and breaking of chemical bonds \cite{energy-transfer}.

The present work presents a possible alleviation of this conceptual barrier as the thermodynamic process and its resultant output can be "\emph{seen}" via the study of an inexpensive Stirling Engine. Through combining the mechanical work of the gas with a more familiar classical mechanical view of work students can gain insight into the transformation of energy. Students also develop and combine multiple competencies including use of electric circuits, signal detection with oscilloscopes and data analysis. The following section will detail the thermodynamic principles of Stirling Engines. Section \ref{section-exp} will provide an overview of the experimental procedure and Section \ref{section-analysis} will present results and analysis to determine the engine's efficiency.  

\end{section}

\begin{section}{Stirling Engine Thermodynamic Principles}
\label{section-thermo}
Stirling Engines are heat engines that operate through the cycling of a work fluid through expansionary and contractionary states to drive pistons that produce useful work. The expansion and contraction of the gas is driven by exposure to hot and cold plates. Through measuring the mechanical work of the internal gas and the useful mechanical work produced, it is possible to calculate the mechanical efficiency of a Stirling Engine.     

\begin{figure}[htp]
    \centering
    \includegraphics[width=10cm]{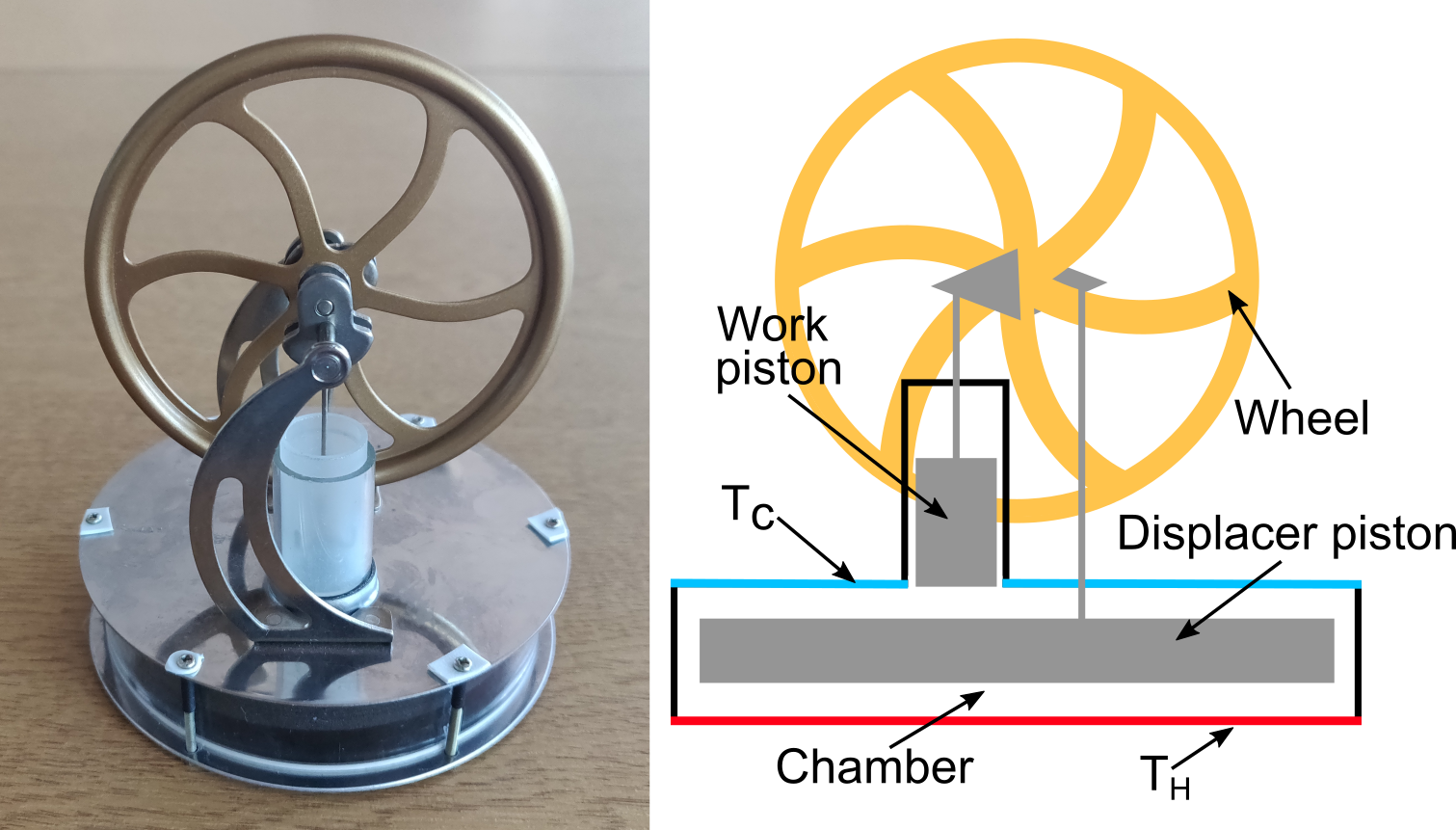}
    \caption{An inexpensive $\beta$ type Stirling Engine is shown (left) together with a schematic showing key components (right). In the latter, the top (cold) and bottom (hot) plates are coloured blue and red respectively. }
    \label{stir-pic-diag}
\end{figure}

The type of Stirling Engines of interest in this study are one chamber $\beta$ type engines with two dissimilar pistons shown in Figure \ref{stir-pic-diag}. Placing the Stirling Engine onto a cup of boiling water brings the bottom plate to temperature $T_{H}$ which heats up the air inside the Stirling Engine chamber causing it to expand. This expansion displaces the pistons and begins the cycle whereby the air cools and contracts when it is shuttled to the cold plate at temperature $T_{C}$. One piston serves as a displacer piston to shuttle the gas between the hot and cold plates and the other piston drives the rotation of the wheel. 

\subsection{Pressure - Volume diagram of a Stirling Engine}
\label{pressure-volume-section}
The development of an analytical model is presented here and is suitable for an early undergraduate physics course. To begin, we can express the thermodynamic cycle of a Stirling Engine in a Pressure - Volume (P-V) diagram. The four corners of the P-V diagram represent the idealized four combinations of pressure and volume that define the borders between the thermodynamic processes involved in one cycle. In reality, the sharp corners of the P-V diagram shown in Figure \ref{p-v} would be rounded but this idealized representation allows the stages to be understood more easily. The four stages of an idealized Stirling Engine are listed below.  

\begin{enumerate}[label=\arabic*)]
    \item Isothermal expansion: The work fluid (air) expands at constant temperature, $T_{H}$, moving the work piston, the temperature remains constant and the pressure reduces. 
    \item Isovolumetric cooling: The temperature of the enclosed air reduces to $T_{C}$ and the pressure drops further. The volume remains constant as the displacer piston moves the gas to the cold side of the engine.
    \item Isothermal compression: The volume of the gas reduces at constant temperature, $T_{C}$ and the pressure increases. Both pistons are pulled inward by the compressing gas displacing the gas to the hot side of the engine.
    \item Isovolumetric heating: The temperature of the gas increases from $T_{C}$ to $T_{H}$ whilst the volume remains constant.
\end{enumerate}

\begin{figure}[htp]
    \centering
    \includegraphics[width=6cm]{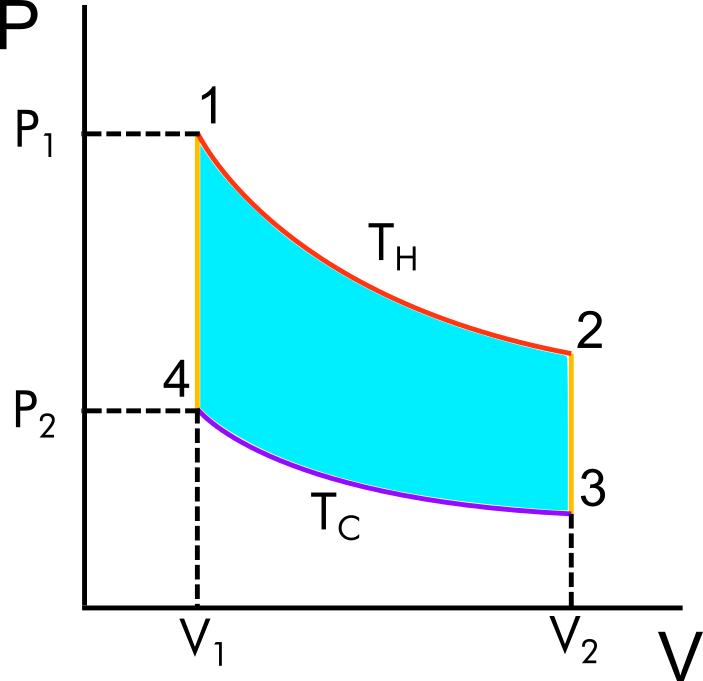}
    \caption{Pressure - Volume diagram for an idealized Stirling Engine.}
    \label{p-v}
\end{figure}

The Stirling Engine shown in Figure 1 can provide an effective way to visualise the stages of the P-V diagram. As the chamber is transparent, students can identify the position of the displacer piston together with the work piston as the engine moves through multiple cycles. In practice however guidance from teachers is typically required for students to fully identify the specific positions of the pistons.

\subsection{Theory}
We are interested in determining the mechanical efficiency, $\eta$ of the Stirling Engine; how much useful work ($W$), is performed by the engine through spinning the wheel compared to the net work done by the gas. Thus, $\eta$ can be expressed as

\begin{equation}
\eta = \frac{W_{wheel}}{W_{gas}}.
\label{eta-eqn}
\end{equation}

The mechanical work done by the Stirling Engine wheel is defined as its rotational kinetic energy:

\begin{equation}
\label{work}
W_{wheel} = KE_{wheel} = \frac{1}{2} I \omega^2.
\end{equation} 

where $I$ is the rotational inertia of the wheel and $\omega$ is the angular velocity of the wheel. We have approximated the wheel as a disk of radius $r$ and mass $M$ with an angular velocity $\omega = 2\pi f$. Equation \eqref{work} can thus be written as:

\begin{equation}
\label{work-updated}
W_{wheel} = \frac{1}{2} \left(\frac{1}{2}Mr^2\right) \omega^2 = \frac{1}{2} \left(\frac{1}{2}Mr^2\right) \left(2\pi f\right)^2 = Mr^2 \pi^2 f^2.
\end{equation}

Next, as depicted in Figure \ref{p-v}, we can express $W_{gas}$ as the heat entering the cycle in steps 4 $\xrightarrow{}$ 1 minus the heat leaving the cycle in steps 2 $\xrightarrow{}$ 3. We will assume that the gas equilibrates with the temperatures $T_{H}$ and $T_{C}$ of the bottom and top plates that are shown in Figure~\ref{stir-pic-diag}. The work done by the gas is expressed through $\int P \cdot dV$. Hence $W_{gas}$ can be expressed as: 

\begin{equation}
W_{gas} = W_{1\rightarrow{}2} + W_{2\rightarrow{}3} + W_{3\rightarrow{}4} + W_{4\rightarrow{}1} = \int P \cdot dV.
\label{overall-p-v-work-eq}
\end{equation}

However, in steps 4 $\xrightarrow{}$ 1 and 2 $\xrightarrow{}$ 3 the gas changes isochorically ($\Delta V = 0$) thus the work is equal to zero:

\begin{equation}
 W_{4\rightarrow{}1} = W_{2\rightarrow{}3} = \int P \cdot dV = 0.
\label{p-v-work-eq}
\end{equation}

In step 1 $\xrightarrow{}$ 2 (3 $\xrightarrow{}$ 4), the pressure and volume change simultaneously during isothermal expansion (compression). Here equation \ref{p-v-work-eq} is $\neq 0$ and we can use the ideal gas law, $PV = nRT$, to derive the work done by the gas. For 1 $\xrightarrow{}$ 2, equation \ref{p-v-work-eq} becomes

\begin{equation}
W_{1\rightarrow{}2} = \int_{V_{1}}^{V_{2}} P \cdot dV = \int_{V_{1}}^{V_{2}} \frac{nRT_{H}}{V} \cdot dV = nRT_{H} \cdot \ln{\frac{V_{2}}{V_{1}}}
\label{pv-expn}
\end{equation}

and in step 3 $\xrightarrow{}$ 4, equation \ref{p-v-work-eq} becomes

\begin{equation}
W_{3\rightarrow{}4} = \int_{V_{2}}^{V_{1}} P \cdot dV = \int_{V_{2}}^{V_{1}} \frac{nRT_{C}}{V} \cdot dV = nRT_{C} \cdot \ln{\frac{V_{1}}{V_{2}}}.
\label{pv-comp}
\end{equation}

The gas inside of the Stirling Engine that that performs the work is termed the work fluid. The ratio of the maximum volume of work fluid, $V_{2}$, to the minimum volume of the work fluid, $V_{1}$, is called the compression ration, $C_R$. In the case of the Stirling Engine we used, shown in Figure \ref{stir-pic-diag}, this is the ratio of the maximum volume of air between the plunger and hot plate during the cycle divided by the maximum volume of air between the plunger and cold plate during the cycle. Using equations \ref{pv-expn} and \ref{pv-comp}, equation \ref{overall-p-v-work-eq} can be written as: 

\begin{equation}
\begin{split}
W_{gas} & = W_{1\rightarrow{}2} + W_{3\rightarrow{}4} \\ 
& = nRT_{H} \cdot \ln{\frac{V_{2}}{V_{1}}} + nRT_{C} \cdot \ln{\frac{V_{1}}{V_{2}}} \\
& = nRT_{H} \cdot \ln{C_R} + nRT_{C} \cdot \ln{\frac{1}{C_R}} \\ 
& = nR\left(T_{H} - T_{C}\right) \cdot \ln{C_R}
\end{split}
\label{q_net-eqn}
\end{equation}

The thermal efficiency can be determined using equation \ref{q_net-eqn} as described in \ref{Appendix-eff}. However, for this student investigation, we are interested in the mechanical efficiency. This can be derived by substituting equations \ref{work-updated} and \ref{q_net-eqn} into equation \ref{eta-eqn}:

\begin{equation}
\label{final-eq}
\eta = \frac{Mr^2 \pi^2 f^2}{nR\left(T_{H} - T_{C}\right) \cdot \ln{C_R}}.
\end{equation}

A final form relevant for the experiment can be written in terms of measured variables of $f$ and $\Delta T = T_{H} - T_{C}$. 
\begin{equation}
\label{ymxc-eq}
f^2 = \frac{n R \cdot \ln{C_R}}{Mr^2 \pi^2} \eta \Delta T.
\end{equation}
Hence, equation \ref{final-eq} can be compared to a linear function $y = mx + c$ in completing a regression analysis to determine the efficiency, $\eta$. This is described further in Section~\ref{section-analysis}.
\end{section}

\begin{section}{Experimental procedure}
\label{section-exp}
This experiment requires the following equipment: a Stirling Engine; a light gate, to measure the engine's rotation; a breadboard with approximately 10 wires and two 100 $\Omega$ resistors; a benchtop power supply; an oscilloscope; two thermocouple probes with tape to attach them; a cup and a kettle to boil water. As shown in Figure \ref{set-up-sch}, the Stirling Engine is mounted on a cup of boiling water that serves as a hot source to create a temperature difference across the chamber of the Stirling Engine. Thermocouples are placed on the top and bottom of the Stirling Engine to measure this temperature difference. A Light Gate is mounted across the wheel to detect its rotation. The light gate is powered by a simple DC circuit shown in Figure \ref{lightgate}. 

\begin{figure}[h]
    \centering
    \includegraphics[width=12cm]{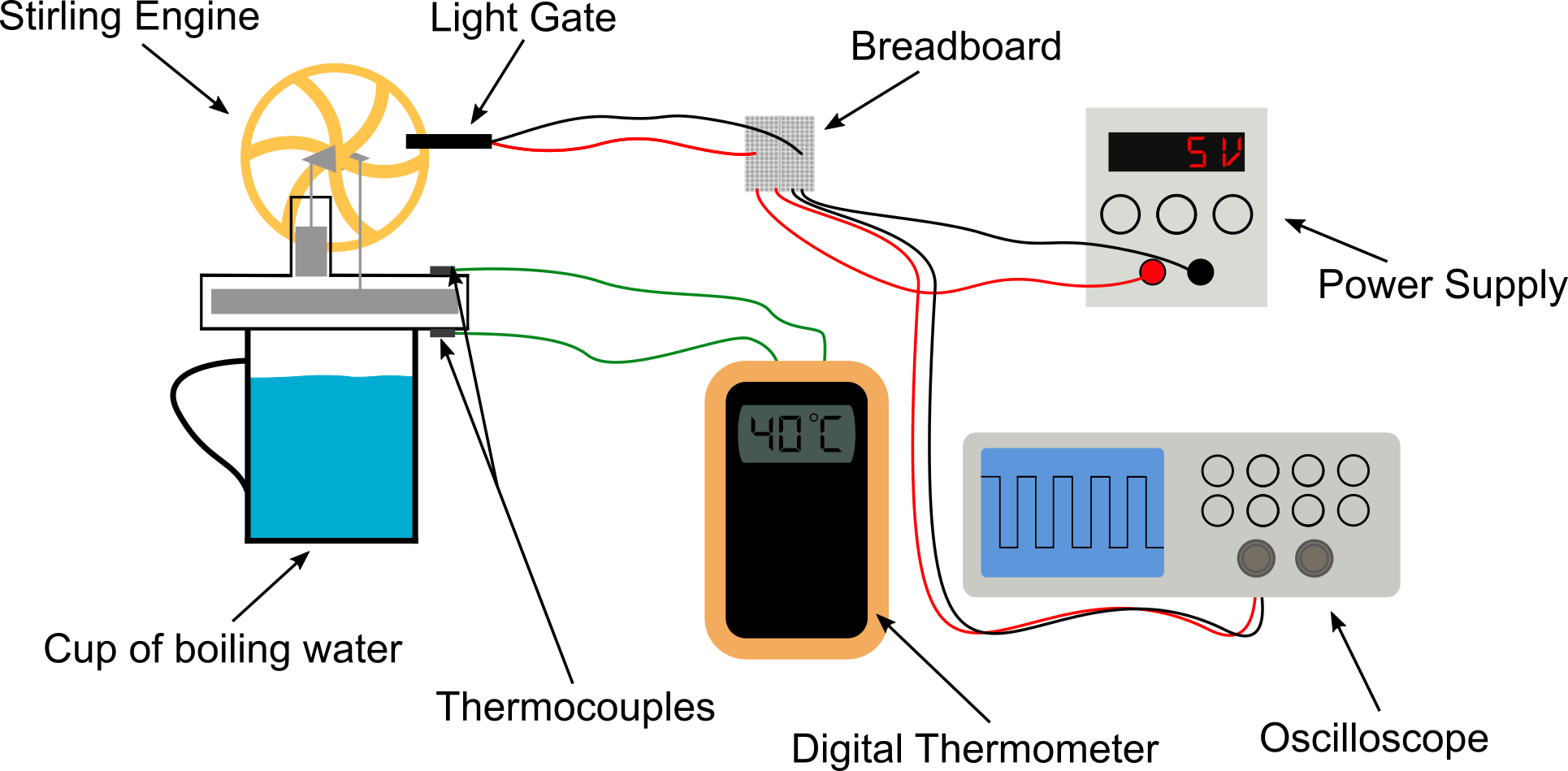}
    \caption{Experimental setup showing key components and their connections. Alternatives to equipment such as the oscilloscope and the digital thermometer could be employed without affecting the experimental outcome.}
    \label{set-up-sch}
\end{figure}
 
For our student's experiment, the Stirling Engines were purchased from a supplier on Amazon.com for £30 each. The wheel of the engine weighed \SI{65.20 \pm 0.01}{\g} and its radius was \SI{4.35 \pm 0.01}{\cm}. Through deconstructing our Stirling Engine and measuring the internal dimensions of the chamber and plunger, we calculated the volume of the internal air and used the density of air to calculate its weight to be \SI{0.1}{\g}. Taking the molar mass of air to be \SI{28.97}{\g\per\mol} we found $n$ in equation \ref{final-eq} to be \SI{3.45}{\milli\mol} of gas. Students determine a value of $C_R$ by measuring the relevant maximum distances of the displacement pistons relative to the hot and cold plates. These values describing the engine are summarised in Table \ref{parameters}. 

\begin{figure}[h]
    \centering
    \includegraphics[width=8cm]{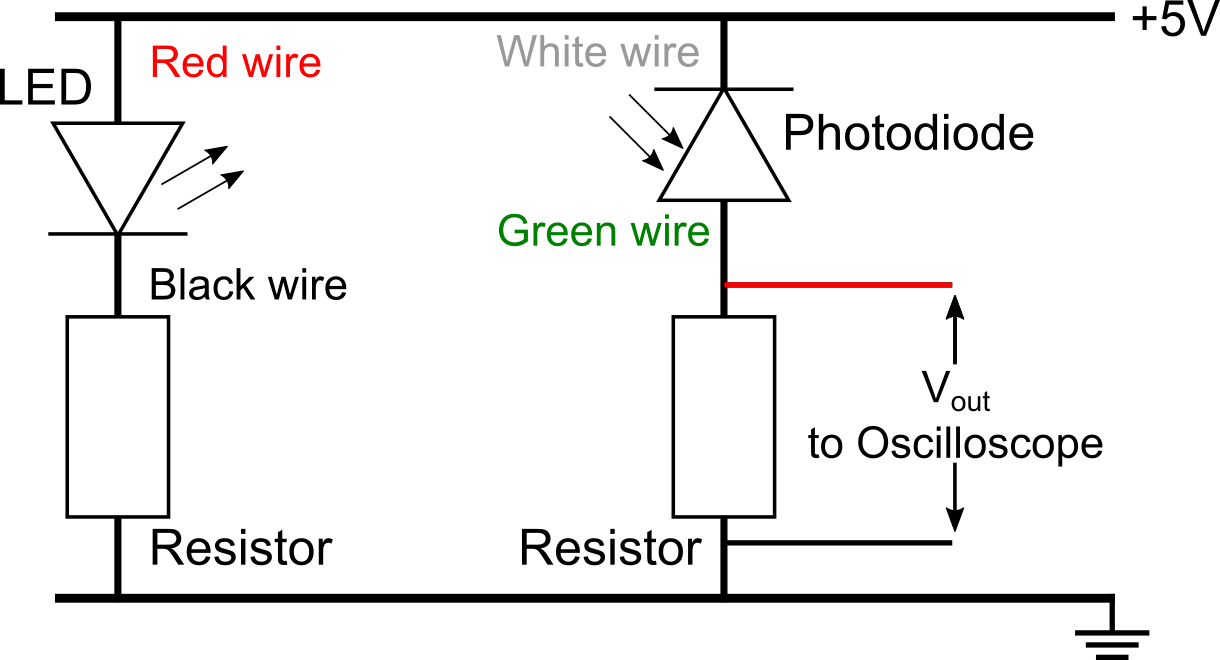}
    \caption{Circuit diagram for the Light Gate breadboard circuit. Note the wire color descriptions were added to aid students as it was found to be a common impediment to correctly wiring the Light Gate.}
    \label{lightgate}
\end{figure}

The experiment is performed through initially filling a cup with boiling water and then immediately placing a room temperature Stirling Engine on top of the cup. Within \SI{90}{\second} the Stirling Engine wheel begins to turn, accelerating rapidly as a temperature differential between the hot and cold plates is now present. Sometimes, a slight push on the wheel is required to overcome its static inertia. The light gate is placed to allow the wheels spokes of the engine to cut across it. When $V_{\mathrm{out}}$ is measured via an oscilloscope, the passage of the spokes through the Light Gate will appear as distinct steps in the waveform. While digital oscilloscopes can measure the frequency of this waveform automatically, students should be encouraged to perform manual measurements of $f$ to build confidence in the data. For example, students can measure the time between six peaks to determine the rotational frequency. As its role is to measure the frequency, the oscilloscope in this experiment could be replaced with a computer with an appropriate I/O device. The separate digital thermometer could be replaced in a similar manner.

As the cup of boiled water cools down, the temperature difference across the chamber of the Stirling Engine reduces and the speed of rotation slows. This creates a decrease in the independent variable: the temperature differences $\Delta T$.  The student's data acquisition will be recording the corresponding rotational frequency, $f$, for a range of $\Delta T$. The subsequent analysis of this data together with equation \ref{ymxc-eq} can be used to calculate the engine efficiency $\eta$. 

\begin{table}
\caption{\label{parameters} Parameters of Stirling Engine used in this experiment.}
\begin{indented}
\item[]\begin{tabular}{@{}lll}
\br
 Parameter         & symbol & value\\
\mr
Mass of wheel & $M$ &  \SI{6.52 \pm 0.01 e-2}{\kilo\gram}      \\
Radius of wheel &  $r$ &  	\SI{4.35 \pm 0.01 e-2}{\meter}     \\
Number of moles of gas   & $n$ &	\SI{3.45 \pm 0.01 e-3}{\mol}  \\
Gas constant & $R$ & \SI{8.31}{\joule\per\mol\per\kelvin} \\
Compression ratio & $C_R$ & $1.3\pm0.3$ \\
\br
\end{tabular}
\end{indented}
\end{table}

\end{section}

\begin{section}{Measurements and analysis}
\label{section-analysis}
As described in section \ref{section-exp}, the variation of rotational frequency with temperature difference comprise the main data for this experiment. As the hot source gradually cools, students can record this data, $f(\Delta T) $, at regular intervals until $\Delta T$ is sufficiently low whereupon the engine stops. Students are encouraged to repeat the experiment a number of times to enable a statistical analysis to be performed. Given the likely imperfections of these engines, together with slight variations in method, this repeat data will serve to demonstrate statistical variation to students. 

Recording repeat data sets can involve two approaches. The first approach is to record $f(\Delta T) $, at distinct values of $\Delta T$. Here students would need to be attentive to record $f$ at the same values of $\Delta T$ during each repeat as the hot source cools, for example at $\Delta T = 60.0$K, 50.0K, 40.0K etc. The difficultly here is trusting that the system will conveniently return to these precise values in subsequent experiments. The second approach is to simply measure $f$ at arbitrary vales of $\Delta T$ and to do at more regular intervals for example at $\Delta T = 60.20$K, 59.31K, 58.10 K etc. Here the measurements could be sampled at a constant rate every few seconds. Each subsequent repeat of the experiment would then follow a similar sample rate without having to match the same $\Delta T$ values. This approach naturally results in larger data sets and the analysis to determine the statistical averages will be more involved.

\begin{figure}[htp]
    \centering
    \includegraphics[width=11cm]{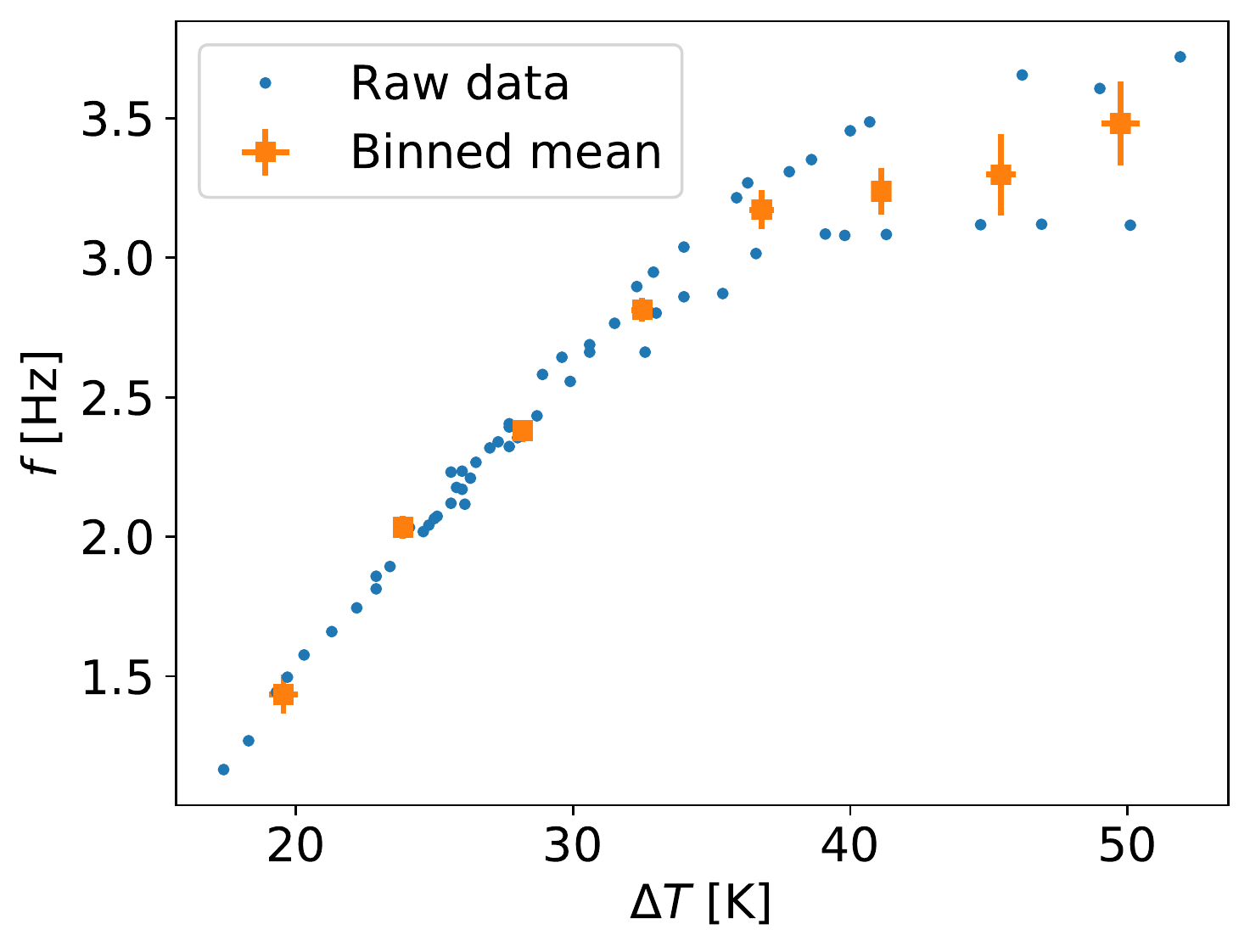}
    \caption{Example student data of rotational frequency, $f$, of the Stirling Engine as a function of temperature difference $\Delta T$. Here the raw data comprises of 3 repeat experiments. The average of the data grouped into 8 bins is also shown together with error bars representing the error on the mean in both variables. Note the clear divergence of measurements of frequency at higher values of $\Delta T$ in the student's data is indicative of a systematic error between repeats.}
    \label{binned}
\end{figure}

\subsection{Binned statistics}\label{binning}
The combining of repeat data at arbitrary values of $\Delta T$ presents an opportunity to teach a useful analysis method called binned statistics. First the repeat data is merged into a single set, this is then combined further into bins by a specific algorithm. Statistics can then be calculated for the contents for each of these $N$ bins. Fortunately, the algorithm for performing this is available to students via the Python programming language. An implementation of this is presented in~\ref{bin-algo}. 

An example result is shown in Figure~\ref{binned}. Here the student recorded raw data for 3 repeats of the experiment. The data was combined into $N=8$ bins across the range of $\Delta T$. The average and standard deviation of the $f(\Delta T)$ values were calculated within each of these bins. The statistical errors shown in Figure~\ref{binned} represent the error on the mean in $f$ and $\Delta T$.  This raw data shows a  divergence of the repeat data at high $\Delta T$ indicating a systematic error in the repeat. This is clearly represented by the larger error bars of the binned data.

\subsection{Regression analysis}
Once the repeat data sets have been combined, students next carry out a regression analysis to determine the engine's efficiency. The theoretical model presented to the students has a dependence of $f^2(\Delta T$) as described by equation \ref{final-eq}. The simplest regression method for students to use here is the least squares approach. Use of this algorithm is standard in most physics courses and its implementation is available in all analysis software.  To perform the analysis, students equate equation~\ref{final-eq} to the linear function $y = mx + c$. The algorithm calculates the values of the gradient, $m$, and the intercept, $c$, which best fit the experimental data. The final step is then to use the gradient value
\begin{equation}
\label{eq-eff}
m = \frac{ \eta n R \cdot \ln{CR}}{\pi^2r^2M}
\end{equation}
together with basic error propagation to determine the engine efficiency $\eta\pm\Delta \eta$.

\begin{figure}[htp]
    \centering
    \includegraphics[width=11cm]{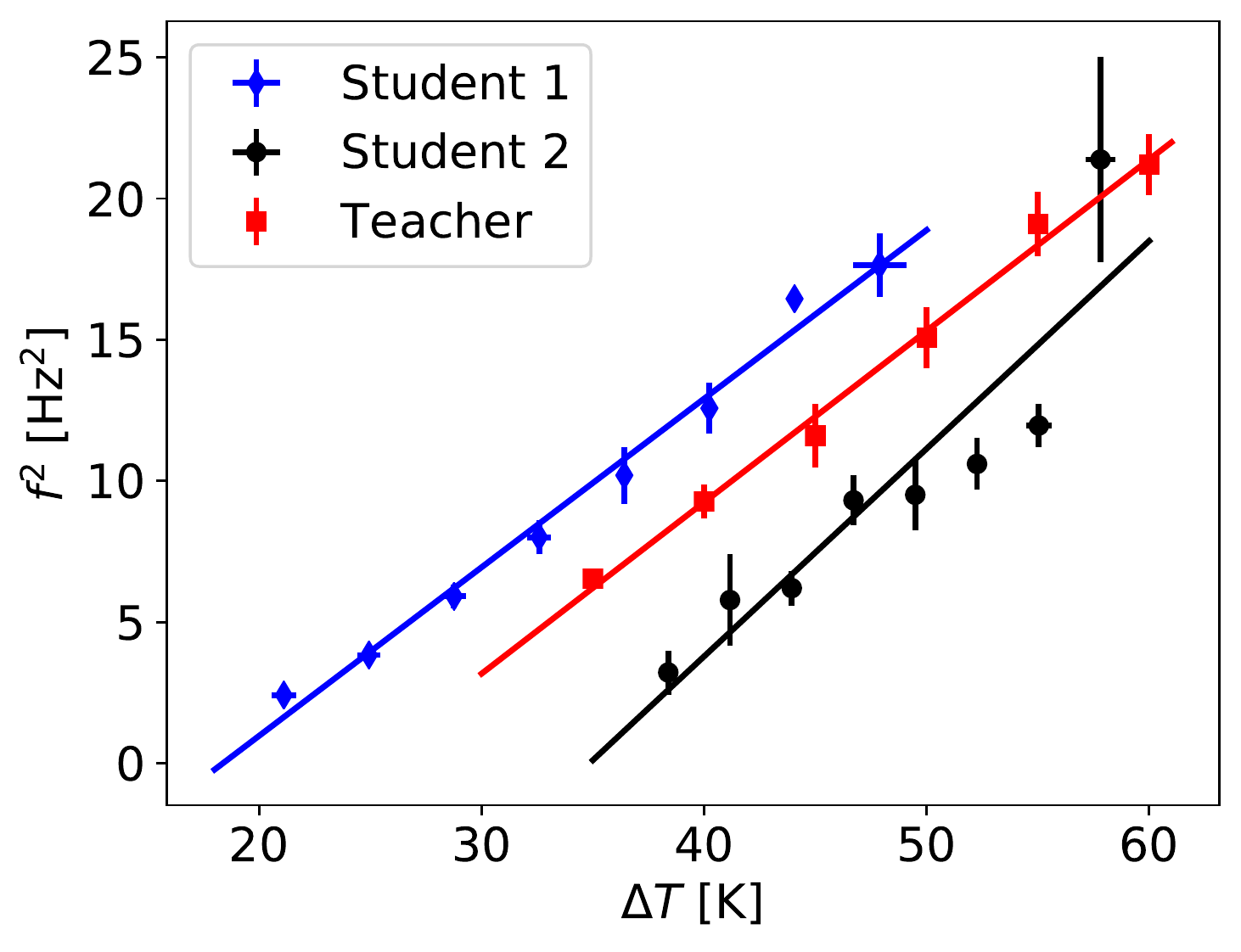}
    \caption{Results from linear regression analysis on experimental data gathered by two students and a teacher. Each includes statistical error bars representing the error on the mean of the repeat measurements. }
    \label{results}
\end{figure}

Representative examples of this regression analysis are shown in Figure~\ref{results}. Here, data sets originally recorded by students has been used as a comparison to data recorded by teaching staff. All three sets of measurements were recorded using individual Stirling Engines of the type described in Section~\ref{intro}. The student repeat data sets were binned using the approached described in Section~\ref{binning} while the teacher data was recorded at discrete values of $\Delta T$ and averaged using several repeat experiments. 

\subsection{Analysis results}
The results of the regression analysis are summarised in Table~\ref{reg-results}. Values of the gradient range from 0.6 $-$ \SI{0.7}{\hertz\squared\per\kelvin} with relative errors ranging from $5-14\%$. The y-intercept values can be used to determine the x-intercept (the minimum value of $\Delta T$ for rotation) which range from 20-40\,K. The goodness of fit, $r^2$, for both Student 1 and Teacher are $>0.9$ but for Student 2, $r^2<0.9$ which is due to the outlier at high $\Delta T$. In the absence of this outlier, the goodness of fit for Student 2 is $r^2=0.9672$ and the corresponding slope is \SI{0.58 \pm 0.05}{\hertz\squared\per\kelvin} .

\begin{table}
\caption{\label{reg-results} Results of the linear regression analysis are shown for the three data sets. Values for gradient, $m$, intercept, $c$, and goodness of fit, $r^2$ are shown. The engine efficiency can be calculated using the gradient result of the regression analysis.  }
\begin{indented}
\item[]\begin{tabular}{@{}llllll}
\br
          & $m$ [\SI{}{\hertz\squared\per\kelvin}] & $c$ [Hz$^2$] & $x (y=0)$ & $r^2$ &  Efficiency {[}\%{]}  \\
\mr
Student 1 & $0.60 \pm 0.03$ & $-11 \pm 1$ & $18\pm2$ & 0.9877 & $10     \pm     2$  \\
Student 2 & $0.7 \pm 0.1$ & $-26\pm 7$ & $40\pm10$ & 0.8245 &  $12      \pm 3 $  \\
Teacher   & $0.61 \pm 0.03$ & $-15 \pm 1$ & $25\pm2$ & 0.9923 & $10      \pm     2 $\\
\br
\end{tabular}
\end{indented}
\end{table}

Finally, the efficiency of the Stirling Engines can be calculated. From Equation 10, we can note that the efficiency will be proportional to the gradient determined from the regression analysis as well as the square of the wheel size and the mass of the wheel. The resulting efficiency values are all approximately 10\%. Given the combination of the uncertainty on the gradient and the compression ratio, the overall relative error on $\eta$ is $20-25\%$. These efficiency results are also summarised in Table~\ref{reg-results}.

For comparison, an ideal machine has a mechanical efficiency of 100\% and a typical internal combustion engine has a mechanical efficiency of around 95\% \cite{mech-eff-engine}. Thus a 10\% mechanical efficiency is very low and this means around 90\% of the potential power output is lost to the surroundings. Students should also be encouraged to make the connection between this mechanical efficiency result and the thermal efficiency. The latter can be generously estimated as 18\% by applying a Carnot model as described in \ref{Appendix-eff}. Hence, a maximum of 18\% of the initial heat energy can be converted into work done by the gas for moving the wheel. Just 10\% of this amount is actually converted to kinetic energy of the rotating wheel thus the overall efficiency is 1.8\%. A typical internal combustion engine found in a car could have a thermal efficiency of 35\% \cite{overall-eff-engine} which using the aforementioned mechanical efficiency gives an overall efficiency of 33.25\%. This is considerably more than 1.8\% yet an internal combustion engine would have a compression ratio at least 10 times as large as the Stirling Engine used in this experiment and would operate at much higher temperatures. 

\end{section}

\begin{section}{Discussion and Conclusion}

The inexpensive Stirling Engines used in this study have been shown to work successfully in an experiment that can be incorporated into a first year undergraduate degree. Even for non-scientists, the sight of the engine beginning to quickly cycle after placement on top of a cup of hot water, then slowing as the water cools, provides a direct visual representation of energy transformation. For students of thermodynamics, the work done by the gas causing the wheel to spin ties the theoretical pressure - volume changes with an intuitive classical mechanics idea of work. Student's conceptual framework of the work done by the gas can be further explored by asking them to identify the processes listed in Section \ref{pressure-volume-section} with the individual strokes of the Stirling Engine's pistons during one full cycle. 

This experiment can also be integrated into wider experimental skills development. For example, prior sessions can be devoted to developing competence with DC circuits, motion sensing and signal measurement via oscilloscopes. The subsequent engine experiment can then incorporate these individually developed skills to build and carry out an experimental investigation. Further skills development including merging repeat data and performing statistical data analysis can follow on from the experimental session. This multi-session approach was the case for the student cohorts who performed this experiment at the author's institution.

In teaching this experiment with two first year undergraduate cohorts, the vast majority complete the experiment within three hours. However, there are some regular issues students encountered with the experiment. A typical problem was the wheel not moving as the plates heat up. This is solved simply by giving the wheel a slight push to overcome the static inertia. Students may also struggle wiring up the Light Gate circuit shown in Figure \ref{lightgate}. We found that explicitly showing the wire colors on the figure used in the lab handout and emphasizing to students to beware of inadvertent contact between circuit components helped with this. 

Occasionally, after some use, the Stirling Engines would seize up and either not cycle or cycle slowly despite an obviously large temperature difference. This can be solved by taking the wheel off and pulling the smaller piston completely out its containment tube. After this, upon reassembly the Stirling Engine performs as normal. We speculate this is due to air gradually being forced out of the central engine chamber as the pistons cycle which alters the pressure inside the chamber. Another cause of slow cycling occurs when the Stirling Engine wheel spins slightly off-axis creating a fishtail motion. The solution here is to alter the axis alignment by wiggling the wheel and bending the support arms inwards to secure the wheel better.  

To improve performance of the Stirling Engines, optional modifications can be added. Lubricant can be applied to the pin bearings and piston rods to ease their movement and thermal insulation can be added to achieve a higher $\Delta T$ across the top and bottom plates. Looking closely at the top plate in the photo on the left of Figure \ref{stir-pic-diag} reveals the thermal insulation modification. Whilst this is a large amount of work for multiple engines, this modification provides a strong performance improvement with peak $\Delta T > \SI{65}{\kelvin}$ and a slower reduction in temperature difference compared to without the modification.

Due to $\Delta T$ being a temperature difference, it is possible to drive the Stirling Engine through a colder $T_{C}$ rather than a hotter $T_{H}$. This can be achieved though setting the Stirling Engine on a Petri dish full of dry ice to cool the bottom plate to around \SI{-70}{\degreeCelsius}. Hence with the top plate at room temperature, a higher maximum $\Delta T \approx \SI{90}{\kelvin}$ can be achieved. However, thermal conduction will also gradually cool the top plate reducing the $\Delta T$ until the wheel slows to a stop. Interestingly, placing the dry ice on the top plate of the Stirling Engine will cause the wheel to spin in the opposite direction as the $T_{C}$ and $T_{H}$ plates are \emph{"flipped"}. However, the advantage of placing the Stirling Engines on top of dry ice is that the teacher can prepare this more easily so the students do not have to handle the dry ice.

To conclude, the thermodynamics experiment described in this paper is a valuable addition to an undergraduate first year physics course. The experiment provides students the opportunity to connect thermodynamic process functions with physical movements of a wheel creating a conceptual framework for work. Students are exposed to oscilloscopes, DC circuits and light gates creating a thorough experience in physics experimentation. Further, the regression analysis and use of binned statistics provides students with a gentle introduction into data analysis.  
\end{section}


\ack
We are grateful to Dr. Stephen Collins and Richard Webb for their technical support  and to Jennifer Bartlett for assisting with the initial conception of this experiment. We also wish to thank the \href{https://www.sheffield.ac.uk/physics/research/shepherd}{ShePHERD} group for contributing to the final proofing of this article. 

\appendix
\section{Thermal efficiency}
\label{Appendix-eff}
The student experiment described in the main paper aims to determine the mechanical efficiency of the Stirling Engine - the useful work divided by the outputted work. However, to determine the thermal efficiency the work done \emph{on} the system must be evaluated. This can be found through transforming equation \ref{p-v-work-eq}, and therefore equation \ref{q_net-eqn}, into 

\begin{equation}
W_{output} = - \int P \cdot dV
= - nR(T_{H} - T_{C}) \cdot \ln{CR}.
\label{work-on}
\end{equation}

The heat inputted into the system comes from the hot plate $Q_{H} = T_{H} \Delta S$. Using the first law of thermodynamics and equation \ref{pv-expn} we find $Q_{H}$ to be 

\begin{equation}
Q_{H} = - nRT_{H} \cdot \ln{CR}.
\label{heat-input}
\end{equation}

Through combining equations \ref{work-on} and \ref{heat-input} the Carnot efficiency can thus be recovered

\begin{equation}
\eta = \frac{W}{Q_{H}} = \frac{-nR(T_{H} - T_{C}) \cdot \ln{CR}}{-nRT_{H} \cdot \ln{CR}} 
= 1 - \frac{T_{C}}{T_{H}} 
\label{carnot}
\end{equation}

The maximum Carnot efficiency, using a $\Delta T = \SI{65}{\kelvin}$ and $T_{H} = \SI{85}{\degreeCelsius}$ is around 18\%. However our simplified view does not take into account the additional terms included in $Q_{H}$ such as regenerative heat loss

\begin{equation}
Q_{r} = MC_{V}(1 - \epsilon_{r})(T_{H} - T_{C})
\label{regenerative}
\end{equation}

Where $M$ is the molar mass of the work fluid, $C_{V}$ the molar specific heat capacity at constant volume of the work fluid and $\epsilon_{r}$ is the regenerator effectiveness. \ref{regenerative} would thus have to be included in the denominator in equation \ref{carnot} meaning we will not reach the Carnot efficiency. For a more detailed breakdown, see \cite{Ahmadi-16}.

\section{Binning algorithm}
\label{bin-algo}
Below is the Python script for binning irregular repeat data used in Figure \ref{binned}:

\begin{lstlisting}[language=Python]
import numpy as np
from scipy.stats import binned_statistic

def average(x, y, nbins):
    """
    Combine data into bins and calculate statistics
    x: independent variable
    y: dependent variable, same size as x
    nbins: number of bins to use
    return: 
    average values for y and x within each bin
    std: standard deviation
    N: number of values used per bin
    """

    # Calculate the mean of the binned data in y
    y_bins, bin_edges, misc = binned_statistic(x, y, statistic="mean", bins=nbins)

    # Calculate the bin centres in x
    x_bins = (bin_edges[:-1] + bin_edges[1:]) / 2

    # Determine std of binned data
    y_std, bin_edges, misc = binned_statistic(x, y, statistic="std", bins=nbins)
    x_std, bin_edges, misc = binned_statistic(x, x, statistic="std", bins=nbins)

    # Determine how many data points N per bin
    N, bin_edges, misc = binned_statistic(x, y, statistic="count", bins=nbins)

    return x_bins, y_bins, x_std, y_std, N
    
\end{lstlisting}

\printbibliography

@article{p-v-pollock,
author = {Pollock,Evan B.  and Thompson,John R.  and Mountcastle,Donald B. },
title = {Student Understanding Of The Physics And Mathematics Of Process Variables In P‐V Diagrams},
journal = {AIP Conference Proceedings},
volume = {951},
number = {1},
pages = {168-171},
year = {2007},
doi = {10.1063/1.2820924},
}

@article{review-mulop,
title = {A Review on Enhancing the Teaching and Learning of Thermodynamics},
journal = {Procedia - Social and Behavioral Sciences},
volume = {56},
pages = {703-712},
year = {2012},
note = {International Conference on Teaching and Learning in Higher Education in conjunction with Regional Conference on Engineering Education and Research in Higher Education},
issn = {1877-0428},
doi = {https://doi.org/10.1016/j.sbspro.2012.09.706},
url = {https://www.sciencedirect.com/science/article/pii/S1877042812041687},
author = {Normah Mulop and Khairiyah Mohd Yusof and Zaidatun Tasir}
}

@article{meltzer-04,
author = {Meltzer,David E. },
title = {Investigation of students’ reasoning regarding heat, work, and the first law of thermodynamics in an introductory calculus-based general physics course},
journal = {American Journal of Physics},
volume = {72},
number = {11},
pages = {1432-1446},
year = {2004},
doi = {10.1119/1.1789161},
URL = {https://doi.org/10.1119/1.1789161},
eprint = {https://doi.org/10.1119/1.1789161}
}

@article{vanRoon-94,
author = { P. H.   van Roon  and  H. F.   van Sprang  and  A. H.   Verdonk },
title = {‘Work’ and ‘Heat’: on a road towards thermodynamics},
journal = {International Journal of Science Education},
volume = {16},
number = {2},
pages = {131-144},
year  = {1994},
publisher = {Routledge},
doi = {10.1080/0950069940160203},

URL = { 
        https://doi.org/10.1080/0950069940160203
},
eprint = { 
        https://doi.org/10.1080/0950069940160203
}
}

@article{Ahmadi-16,
    author = {Ahmadi, Mohammad Hossein and Ahmadi, Mohammad Ali and Mehrpooya, Mehdi},
    title = "{Investigation of the effect of design parameters on power output and thermal efficiency of a Stirling engine by thermodynamic analysis}",
    journal = {International Journal of Low-Carbon Technologies},
    volume = {11},
    number = {2},
    pages = {141-156},
    year = {2016},
    month = {05},
    issn = {1748-1317},
    doi = {10.1093/ijlct/ctu030},
    url = {https://doi.org/10.1093/ijlct/ctu030},
    eprint = {https://academic.oup.com/ijlct/article-pdf/11/2/141/6766247/ctu030.pdf},
}

@article{mech-eff-engine,
    author = {ElBahloul, Mostafa A. and Aziz, ELsayed S. and Chassapis, Constantin},
    title = "{Mechanical efficiency prediction methodology of the hypocycloid gear mechanism for internal combustion engine application}",
    journal = {International Journal on Interactive Design and Manufacturing},
    volume = {13},
    pages = {221-233},
    year = {2019},
    month = {03},
    issn = {1955-2505},
    doi = {10.1007/s12008-018-0508-2},
    url = {https://doi.org/10.1007/s12008-018-0508-2},
}

@article{overall-eff-engine,
author = {Jerald A Caton},
title ={Maximum efficiencies for internal combustion engines: Thermodynamic limitations},
journal = {International Journal of Engine Research},
volume = {19},
number = {10},
pages = {1005-1023},
year = {2018},
doi = {10.1177/1468087417737700},

URL = { 
        https://doi.org/10.1177/1468087417737700
    
},
eprint = { 
        https://doi.org/10.1177/1468087417737700
    
}
}

@article{energy-transfer,
author = {Macrie-Shuck, Michael and Talanquer, Vicente},
title = {Exploring Students’ Explanations of Energy Transfer and Transformation},
journal = {Journal of Chemical Education},
volume = {97},
number = {12},
pages = {4225-4234},
year = {2020},
doi = {10.1021/acs.jchemed.0c00984},

URL = { 
        https://doi.org/10.1021/acs.jchemed.0c00984
    
},
eprint = { 
        https://doi.org/10.1021/acs.jchemed.0c00984
    
}

}

@article{energy-entropy-concept,
	doi = {10.1088/1361-6552/ab4de6},
	url = {https://doi.org/10.1088/1361-6552/ab4de6},
	year = 2019,
	month = {nov},
	publisher = {{IOP} Publishing},
	volume = {55},
	number = {1},
	pages = {015005},
	author = {Guobin Wu and Amy Yimin Wu},
	title = {A new perspective of how to understand entropy in thermodynamics},
	journal = {Physics Education}
}

@article{breath-engine,
author = {Lipscombe,Trevor C.  and Mungan,Carl E. },
title = {Breathtaking Physics: Human Respiration as a Heat Engine},
journal = {The Physics Teacher},
volume = {58},
number = {3},
pages = {150-151},
year = {2020},
doi = {10.1119/1.5145400},

URL = { 
        https://doi.org/10.1119/1.5145400
    
},
eprint = { 
        https://doi.org/10.1119/1.5145400
    
}

}

@article{sunray-concept,
	doi = {10.1088/1361-6404/abce1e},
	url = {https://doi.org/10.1088/1361-6404/abce1e},
	year = 2021,
	month = {mar},
	publisher = {{IOP} Publishing},
	volume = {42},
	number = {3},
	pages = {035101},
	author = {Joon-Hwi Kim and Juno Nam},
	title = {Thermodynamic identities with sunray diagrams},
	journal = {European Journal of Physics}
}

\end{document}